\documentclass[reprint,aps,ams,amsmath,prx,twocolumn,longbibliography,superscriptaddress]{revtex4-1}
\usepackage{graphics}
\usepackage{graphicx}
\usepackage{epstopdf,epsfig}
\usepackage{amsmath}
\usepackage{amssymb}
\usepackage{amsfonts}
\usepackage{mathrsfs}
\usepackage{bm}
\usepackage{color}
\usepackage{bbm}
\usepackage{hyperref}
\usepackage[normalem]{ulem}
\usepackage{comment}
\usepackage{soul}

\DeclareMathOperator{\sgn}{sgn}
\newcommand{\be}{\begin{equation}}
\newcommand{\ee}{\end{equation}}
\def\rr#1{(\ref{#1})}

\begin{document}

\title{Electromotive entrainment of charge and heat currents in graphene}

\author{E. Kirkinis}
\affiliation{Center for Computation and Theory of Soft Materials, Robert R. McCormick School of Engineering and Applied Science, Northwestern University, Evanston IL 60208 USA}

\author{A. Levchenko}
\affiliation{Department of Physics, University of Wisconsin-Madison, Madison, Wisconsin 53706, USA}

\author{A. V. Andreev}
\affiliation{Department of Physics, University of Washington, Seattle, Washington 98195, USA}

\date{\today}

\begin{abstract}
We develop a hydrodynamic theory of charge and heat currents induced by traveling waves, such as surface acoustic waves, in graphene devices near charge neutrality. The currents depend on the intrinsic conductivity and viscosity of the electron liquid, the disorder strength, and the geometry of the device. We obtain analytic expressions for the heat and charge currents to second order in the wave amplitude for Hall-bar devices. At charge neutrality and in the absence of DC bias, the heat content is entrained by the wave in the absence of net charge transfer. At the same time, device conductance is enhanced by the wave. Away from charge neutrality, the transport charge current induced by the wave arises in the absence of a DC bias.
\end{abstract}

\maketitle

\section{Introduction}

At finite temperature, coupling of electrons to external perturbations  creates modulation of both the charge and entropy density of the electron liquid. In situations where such perturbations have a form of a traveling wave, a net transport current of both charge and heat is induced in the system.  Such periodic modulations can be efficiently created in 1D and 2D systems by surface acoustic waves (SAW).

At low temperatures and sufficiently strong drive,   the transport charge current can be quantized under certain conditions. Examples include the Fr\"ohlich current induced by a sliding CDW in quasi-one-dimensional metals~\cite{Frohlich1954}, and the quantum Thouless pumping current~\cite{Thouless1983} in periodic 1D systems.
Quantized charge current induced by SAW has been measured in carbon nanotubes~\cite{Talyanskii2001,Leek2005}.

In the generic situation of nonzero temperature, the transport current of charge and heat induced by the traveling perturbation is not quantized. Its evaluation requires consideration of the evolution of both charge and entropy density. Motivated by applications to graphene systems
\cite{Miseikis2012,Bandhu2013,Santos2018,Pollanen2018,Nichlos-PhDThesis,Ganichev2022,Ramshaw2023,Zhang2025}, in the present article we develop a theory of wave-induced charge and heat transport in the hydrodynamic regime in the vicinity of a charge neutrality point.

The mechanism of entrainment of charge and heat in this case differs from the well known acoustic streaming \cite{Rayleigh1884,Schlichting1932,Landau1987,Hui2021} and acoustoelectric effect \cite{Parmenter1953,Weinreich1957,Gurevich1963}. Since experimentally, the traveling wave potential is typically induced by SAW propagating in a piezoelectric substrate,  we will refer to the potential $U$ as the SAW potential.
The physical mechanism of induced charge and heat flow may be qualitatively described as follows. The electric potential $U$ induced by the traveling wave in the 2D electron channel, produces an electron density modulation $\delta n$, that is linear in $U$ and slightly out of phase with the traveling wave. The coupling of the density modulation  $\delta n$ to $U$ exerts an average drag force on the electron liquid. This induces transport of heat and charge currents in the channel.
The magnitude of the transport current is determined by the balance of the drag force exerted by the traveling wave on the electron density modulation, and the rate of momentum relaxation of the electron liquid. The latter occurs due to disorder in the system or, in the hydrodynamic regime, by momentum outflow to the system boundary via viscous stresses. We consider the crossover between the disorder-dominated and hydrodynamic regime of momentum relaxation. In the hydrodynamic regime, the density of the induced transport current increases with the channel width (or equivalently, the net transport current is superextensive in the channel width).

We show that in graphene devices near charge neutrality, the amplitude of the charge modulation induced by the SAW is controlled by the intrinsic conductivity. As a result, the wave-induced enhancement of the conductivity is proportional to the square of the intrinsic conductivity, while the entrainment currents of charge and heat are linear in it. We further show that at charge neutrality and zero DC electric field bias, the electric current vanishes but the entropy current is non-zero. Thus, similarly to the slow pumping regime~\cite{Andreev2022}, heat entrainment by SAW proceeds in the absence of charge transfer.

In the absence of a DC electric field, the transport charge current arises only at nonzero electron density. In the long-wavelength limit, this current scales quadratically with the SAW  amplitude and linearly with the electron density.

The presentation below is organized as follows. In section \ref{sec: gov} we formulate the general hydrodynamic problem, which we then specialize to the case of a Hall-bar geometry. The wave gives rise to current and conductivity perturbations that are developed in section \ref{sec: DC}. We consider the effects of long-range disorder at finite frequencies of the external excitation in section \ref{sec: disorder}. Weak disorder or narrow channels recover the clean system results, while strong disorder or wide channels lead to a vanishingly small electric current. Detailed calculations are delegated to a number of Appendices \ref{app:velocity}--\ref{sec: validity}.

\begin{figure}
\vspace{0pt}
\centering
\includegraphics[height=2.5in,width=4.4in]{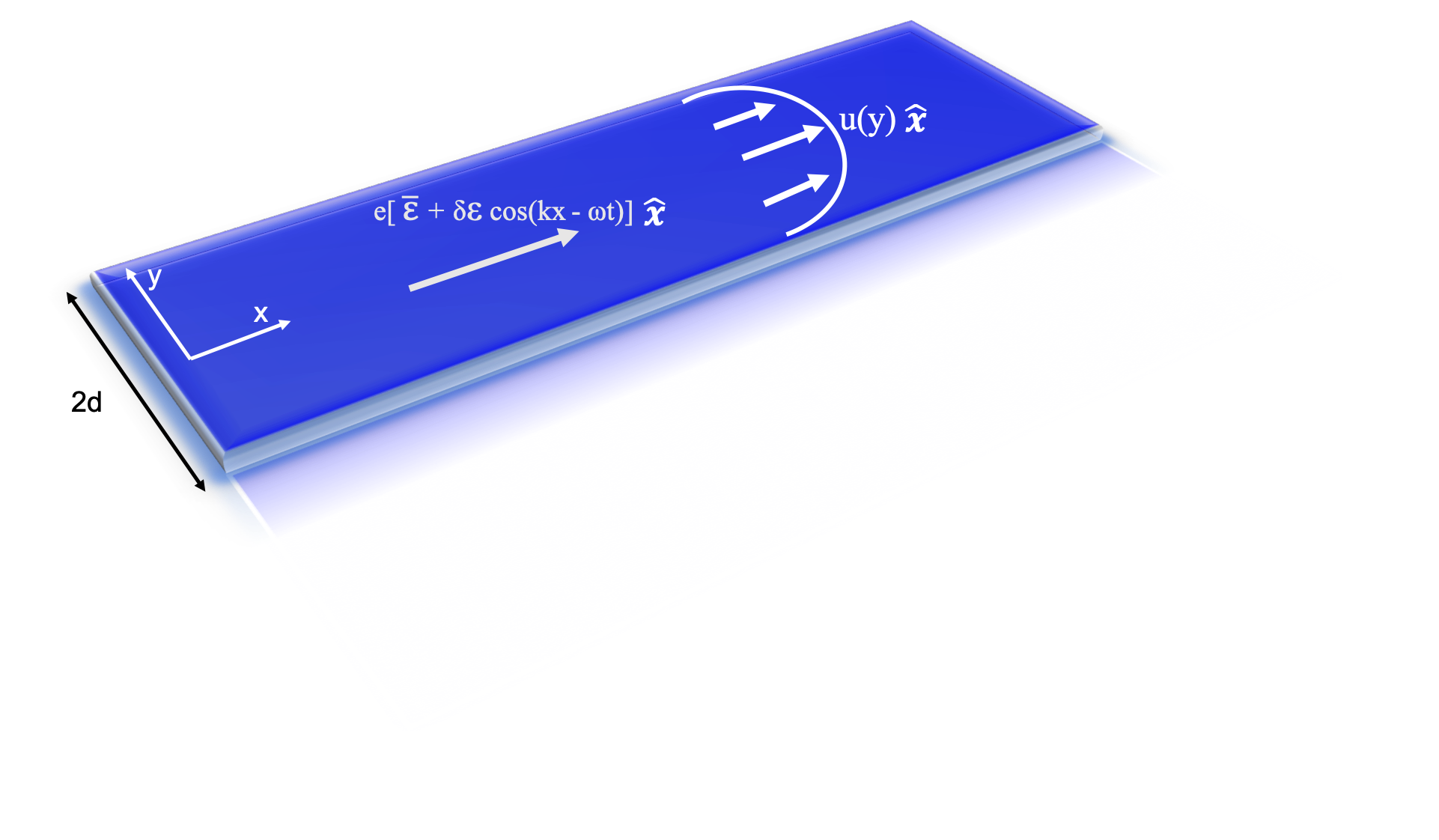}
\vspace{-70pt}
\caption{A Hall-bar configuration subjected to a traveling-wave EMF, $e\delta\mathcal{E}\cos(kx - \omega t)$, superposed on a uniform field $\bar{\mathcal{E}}$ in the $\hat{\mathbf{x}}$ direction, induces (at charge neutrality) a traveling-wave density perturbation $\delta n \cos(kx - \omega t)$, which is itself superposed on a uniform background density $\bar{n}$. This, in turn, gives rise to a DC electric current that enhances the conductivity and entrains both the electron liquid and its heat content.
\label{hall_bar2}}
\vspace{-0pt}
\end{figure}


\section{Wave-induced currents in graphene Hall-bar devices}\label{sec: gov}

We consider a two-dimensional electron system in a Hall-bar geometry, depicted schematically in Fig.~\ref{hall_bar2}, which is perturbed by a traveling wave of external electric potential $U(x- ct)$, with $c$ being the wave velocity.  We assume that the system operates in the hydrodynamic regime, which can be realized within a range of temperatures and sample purities where the momentum relaxation length due to electron–impurity and electron–phonon scattering exceeds the relaxation length associated with momentum-conserving electron–electron collisions \cite{Gurzhi1968,Gao2010,Andreev2011}. The current interest in this transport regime has been stimulated by significant progress in the fabrication of high-mobility, low-density semiconductor heterostructures, as well as boron nitride–encapsulated graphene devices, which exhibit unconventional hydrodynamic effects; see Refs.~\cite{LucasFong2018,Levchenko2020,Narozhny2022} for recent reviews.
The specific applicability conditions for the hydrodynamic limit of electron transport in monolayer and bilayer graphene devices were considered theoretically in Ref. \cite{Adam2018}.

In graphene, the charge and entropy currents can be expressed in terms of the hydrodynamic velocity $\bm{u}$, electromotive force (EMF) $\bm{\mathcal{E}}$, and temperature gradients as follows
\begin{subequations}\label{current1}
\begin{align}
&\mathbf{j}_e = en\bm{u} + \sigma \bm{\mathcal{E}} - \frac{\gamma}{T} e \bm{\nabla} T, \\
&\mathbf{j}_s = s\bm{u} -\frac{\kappa}{T} \bm{\nabla} T +  \frac{\gamma}{T} e \bm{\mathcal{E}}.
\end{align}
\end{subequations}
Here $n$ and $s$ are, respectively, particle and entropy densities,  while $\sigma$ and $\kappa$ are the intrinsic electrical and thermal conductivities of the electron liquid.
Finally, $\gamma$ denotes the thermoelectric coefficient of the electron liquid. Near charge neutrality, where $n \ll s$, this coefficient is small and will be neglected in the analysis below.

The traveling wave potential $U$ induces a spatial modulation of the electron density. Therefore, the EMF is given by
\begin{align}
    \label{eq:EMF_def}
    e\bm{\mathcal{E}} = - \bm{\nabla} (\mu + U + e\phi),
\end{align}
where $\mu$ is the change of the chemical potential, and $\phi$ the change of electric potential caused by the electron density modulation.

The currents from Eq. \rr{current1} satisfy the continuity equations
\be\label{cons1}
\partial_t en = -\nabla\cdot \mathbf{j}_e , \quad \partial_t s = -\nabla \cdot \mathbf{j}_s+\varsigma,
\ee
where $\varsigma$ denotes the local rate of entropy production due to
electron-electron collisions.

At temporal and spatial scales large compared to their microscopic counterparts, the electron liquid flow can be described by the Navier-Stokes equations
\be \label{NS}
\rho(\partial_t+\bm{u}\cdot\bm{\nabla})\bm{u} =  \eta \nabla^2 \bm{u} + \zeta \bm{\nabla}(\bm{\nabla}\cdot\bm{u}) + e n \bm{\mathcal{E}}-s \bm{\nabla} T,
\ee
where $\rho$, $\eta$, and $\zeta$ are the electron liquid mass density, shear, and bulk viscosities, respectively. In \rr{NS} we employed the thermodynamic identity $ dP = n d\mu + s dT$ to eliminate the pressure $P$.

In the following analysis, the liquid is assumed to satisfy no-slip boundary conditions at the lateral walls of the Hall bar located at $y=\pm d$; see Fig. \ref{hall_bar2} for the proposed device geometry. This is not an overly restrictive assumption since, quite generally, the slip length corresponding to more general boundary conditions is of the order of the electron-electron mean free path \cite{Kiselev2019}, which is much smaller than the width of the Hall bar channel. Therefore, it constitutes only a small boundary effect in the hydrodynamic limit. We also recall that, at charge neutrality, monolayer graphene is well described by a (nearly) scale-invariant, conformal electron fluid. In such systems, the bulk viscosity is strongly constrained and vanishes, $\zeta\to0$, to leading order \cite{Principi2016}. It is also negligibly small for bilayer graphene.

For the Hall-bar geometry of Fig. \ref{hall_bar2}, we consider a traveling wave perturbation in the $\hat{\mathbf{x}}$ direction of real amplitude $\delta \mathcal{E}$ superposed on a uniform bias field $\bar{\mathcal{E}}$
\be\label{E}
\mathcal{E} (x,t)=\bar{\mathcal{E}} +  \delta\mathcal{E} \cos(kx-\omega t)
\ee
with $\omega>0$ and $c=\omega/|k|$ the phase velocity, so that the wave travels in the direction $\sgn k$, giving rise to commensurate (complex) density and entropy perturbations $\delta n$ and $\delta s$
\be\label{ns}
\begin{pmatrix}
    n(x,t)\\
    s(x,t)
\end{pmatrix}
=
\begin{pmatrix}
   \bar{n}\\
\bar{s}
\end{pmatrix}
+
\frac{1}{2} \left[
\begin{pmatrix}
   \delta n\\
\delta s
\end{pmatrix}
e^{i(kx-\omega t)}
+ c.c.\right]
\ee
superposed on their respective uniform fields $\bar{n}$ and $\bar{s}$. For weak potential we have $\delta \mathcal{E} = - \alpha \partial_x U$, where the coefficient $\alpha$ must be determined by the Poisson equation \footnote{The relation between the EMF and the external potential of the wave is further discussed in Appendix \ref{sec: validity}.}.
We also replace the temperature gradient with the thermodynamic equality $\partial_xT = \frac{T}{c_p} \partial_x s$, where $c_p$ is the specific heat capacity at constant pressure.

Considering the form acquired by the body force in the momentum equation
\rr{NS} (see Eqs. \rr{enE0} and \rr{sdxT} in Appendix \ref{app:velocity}), subject to \rr{E} and \rr{ns}, leads
the velocity field   $\bm{u}(x,y,t) = u(x,y,t) \hat{\mathbf{x}}$ to become
\begin{align} \label{u}
u(x,y,t) = \bar{u}(y) + \nonumber
\frac{1}{2}\left[ \right.& \left. u_0(y)+u_1(y)e^{i(kx-\omega t)} \right. \\
+&\left.u_2(y)e^{2i(kx-\omega t)}  +c.c.\right].
\end{align}
The steady component $\bar{u}$ gives rise to all the linear-response hydrodynamic thermoelectric effects subject to a time-independent uniform field $\bar{\mathcal{E}}$, discussed in Ref. \cite{Li2022}. On the other hand, $u_0, u_1$ and $u_2$ are (complex-valued) nonlinear perturbations induced by the wave excitation and depend on its frequency and wavenumber. These perturbations are responsible for the induced entrainment effects. Throughout, the density and entropy perturbations are treated as uniform across the channel and the velocities entering the conservation laws are replaced by their width averages (Appendix~\ref{sec: perturbations}); this closure is accurate for $kd\lesssim1$.

The perturbation series is organized as follows. We expand in the wave amplitude $\delta\mathcal{E}$ to second order and in the bias $\bar{\mathcal{E}}$ to first order. The density and entropy perturbations $\delta n$, $\delta s$ in Eq.~\eqref{ns} are of first order in $\delta\mathcal{E}$ (they also contain terms of order $\delta\mathcal{E}\bar{\mathcal{E}}$); their second-order parts, at zero wavenumber and at $2k$, are not needed: the former vanishes for a gated channel at fixed electrochemical potential, and the latter would enter the DC currents only at third order. The velocity, on the other hand, must be kept to second order, because the DC currents, Eqs.~\eqref{dIe-def} and \eqref{dIs} below, are themselves of second order: they contain the product $\delta n^*u_1$ of two first-order quantities and the product $\bar{n}u_0$ of the zeroth-order density with the second-order rectified velocity $u_0\propto\delta n\,\delta\mathcal{E}$. The second harmonic $u_2$ is listed for completeness and does not contribute to the DC currents at this order. Accordingly, the first harmonic of the conservation laws \eqref{cons1} determines $\delta n$ and $\delta s$ (Appendix~\ref{sec: perturbations}), while the zeroth and first harmonics of the momentum equation \eqref{NS} determine $u_0$ and $u_1$ (Appendix~\ref{app:velocity}).

In what follows, we show that, in the absence of the bias field $\bar{\mathcal{E}}$, the traveling wave induces no charge current up to quadratic order in the wave amplitude $\delta\mathcal{E}$. We work to linear order in the DC bias and up to quadratic order in the wave amplitude. For this reason the explicit form of the entropy production rate $\varsigma$ in Eq.~\eqref{cons1} is not required: at $\bar{\mathcal{E}}=0$ it has no component at the wave frequency, and in the presence of the bias its first-harmonic (Joule) part, $\propto\sigma\bar{\mathcal{E}}\delta\mathcal{E}/T$, is smaller than the convective source of $\delta s$ in Eq.~\eqref{cds} by the factor $\sim(c/v)(\lambda_T/d)/kd\ll1$, where $v$ is the band velocity and $\lambda_T=\hbar v/T$ is the thermal de Broglie length.

\section{\label{sec: DC}Clean systems with uniform density}

In linear order in the DC bias field $\bar{\mathcal{E}}$, the DC transport electric current is given by
\be \label{IDC0}
I_e =\bar{I}_e + \delta I_e
\ee
where $\bar{I}_e$ is the current in the absence of the wave~\cite{Li2022}, cf. Appendix \ref{sec: currents}, and
\be \label{dIe-def}
\delta I_e =\Re  \int_{-d}^d e \left\{ \bar{n}u_0(y) + \frac{\delta n^*}{2} u_1(y)  \right\} dy,
\ee
is the nonlinear perturbation induced by the traveling wave field. The latter gives rise to a wave-induced conductivity correction $\delta \sigma$ such that
\be \label{dIe}
\delta I_e = 2d \delta \sigma \bar{\mathcal{E}}
\ee
whose form we derive below.

At $\bar{n} \equiv 0$, with vanishing thermoelectric coefficient $\gamma$ and temporarily leaving $\bar{\mathcal{E}}$ nonzero, the density and entropy perturbations become
\begin{subequations}\label{dnds}
\begin{align}
\delta n & = \frac{k \sigma  \delta \mathcal{E}}{e \omega}, \\
\delta s & = \frac{\bar{s}\mathscr{G}(Kd)d^{2} k^{2} \delta \mathcal{E} \bar{\mathcal{E}} \sigma  \mathit{c_p}}{ \omega \left(i\mathscr{G}(Kd) T \,d^{2} k^{2} \bar{s}^{2}+3 \,ik^{2} \eta  \kappa +3 \mathit{c_p} \eta \omega \right)}
\end{align}
\end{subequations}
by solution of the conservation laws \rr{cons1}, cf. Appendix \ref{sec: perturbations}. Note that $\delta s$ vanishes when $\bar{\mathcal{E}}$ does. The complex valued structure factor
\be \label{G0}
\mathscr{G}(Kd)\equiv\frac{3}{(Kd)^3}\left[\tan Kd - Kd \right]
\ee
is a universal crossover function that determines the dependence of magnitude of the transport current on the sample width, the wave parameters and the kinematic viscosity (cf. Fig. \ref{structure_factor} and Appendix \ref{sec: structure}). The Stokes wavenumber renormalized by the wave excitation wavenumber $k$ is given by
\begin{equation}
\label{eq:K_def}
K = \sqrt{\frac{i\omega}{\nu} - k^2},
\end{equation}
where $\nu = \eta/\rho$ is the kinematic viscosity of electron liquid.

\begin{figure}
\vspace{5pt}
\begin{center}
\includegraphics[height=2.8in,width=3.6in]{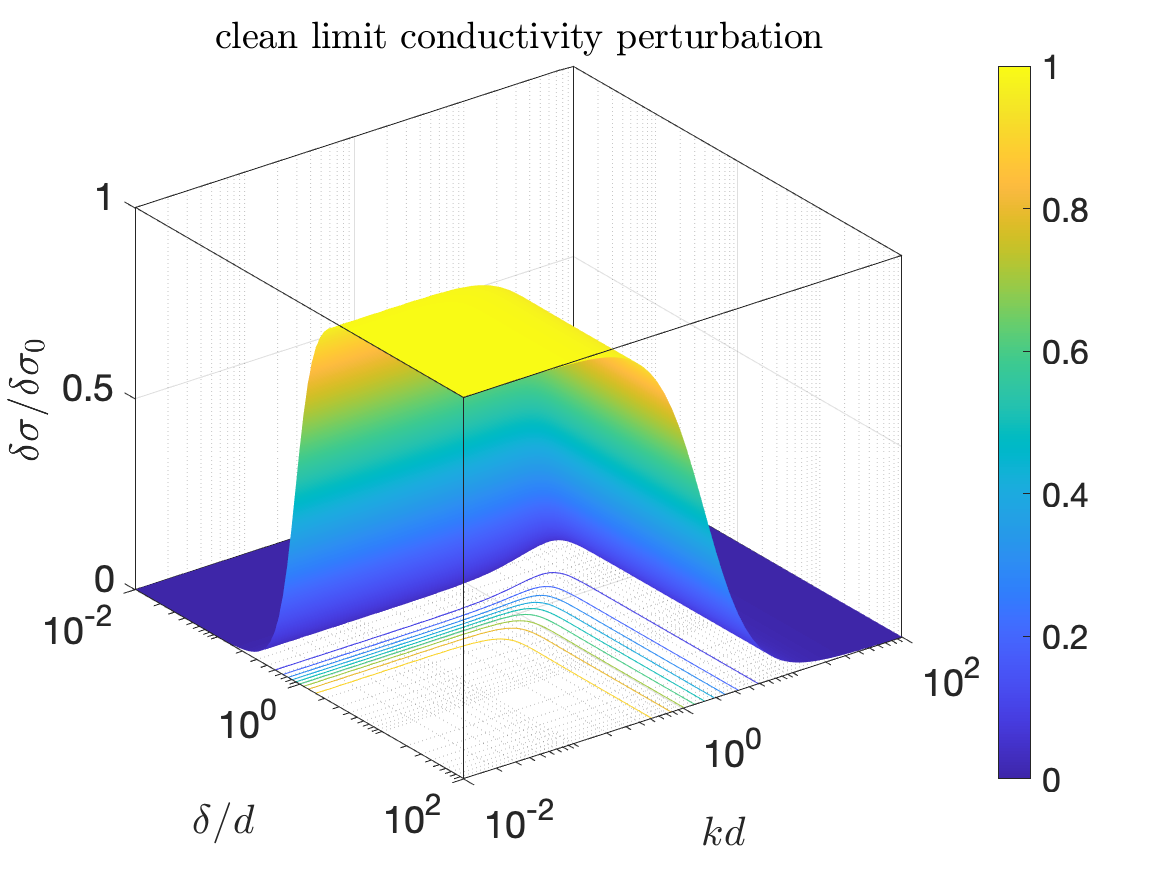}
\vspace{-0pt}
\end{center}
\caption{The wave-induced conductivity enhancement, $\delta\sigma$ defined in Eq.~\rr{dsigma}, is plotted as a function of the dimensionless parameters $kd$ and $\delta/d$. The overall conductivity scale is normalized to the unit $\delta\sigma_0 \equiv \sigma^2 \delta\mathcal{E}^2 d^2 / (6 \eta c^2)$.
\label{deltasigma}}
\vspace{-0pt}
\end{figure}

By considering the velocity fields \rr{uav}, the conductivity perturbation $\delta \sigma$ appearing in the current \rr{dIe} can be presented in the form
\begin{align} \label{dsigma}
&\delta\sigma = \frac{  \sigma^2(\delta \mathcal{E})^2 d^2  }{6\eta c ^2}\times \nonumber \\
&\Re\left[ \mathscr{G}(Kd)- \frac{ \mathscr{G}^2(Kd)  }{ \mathscr{G}(Kd) +  \frac{3\eta  \kappa}{T(\bar{s}d)^2} - 3 i  \frac{c_p \eta c}{Td \bar{s}^2}\frac{1}{kd} } \right].
\end{align}
The characteristic behavior of the induced enhancement of the conductivity mainly depends on a relationship between two dimensionless parameters of the model, $cd/\nu$ and $kd$, and their absolute magnitudes, where $c=\omega/|k|$ is the phase velocity of the wave. For example, if $c\ll\nu k$ we find to the leading order
\be \label{dsigma-c-ll}
\delta \sigma \simeq  \frac{  \sigma^2(\delta \mathcal{E})^2 d^2  }{6\eta c ^2}
\left\{\begin{array}{cc}1 & kd\ll1\\ \frac{3}{(kd)^2}& kd\gg1\end{array}\right..
\ee
Similarly, if $c\gg\nu k$ then asymptotic limits become
\be \label{dsigma-c-gg}
\delta \sigma \simeq  \frac{  \sigma^2(\delta \mathcal{E})^2 d^2  }{6\eta c ^2}
\left\{\begin{array}{cc}1 & d/\delta\ll1\\ \frac{3\delta^3}{4d^3}& d/\delta\gg1\end{array}\right..
\ee
where $\delta \equiv (2\nu/\omega)^{1/2}$ is the penetration depth.
In Fig.~\ref{deltasigma}, we display the characteristic behavior of the conductivity correction \rr{dsigma}, as a function of the dimensionless parameters $\delta/d$ and $kd$. It can be easily checked that, in the parameter regime shown in Fig.~\ref{deltasigma}, the conductivity enhancement is dominated by the first term in the brackets of Eq. \eqref{dsigma}.

In complete analogy, the net entropy current \rr{cons1}, averaged over the period of oscillation,
\be \label{Is2}
I_s = \bar{I}_s + \delta I_s,
\ee
is composed of a linear term
$\bar{I}_s$ as was determined in Ref.~\cite{Li2022}, and a nonlinear perturbation
\be \label{dIs}
\delta I_s =\Re  \int_{-d}^d  \left\{ \bar{s}u_0(y) + \frac{\delta s^*}{2} u_1(y)  \right\} dy,
\ee
induced by the traveling wave field. By examining the velocity fields (see Eq.~\rr{uav} in Appendix \ref{app:velocity}), one can readily extract the leading-order asymptote in the long-wavelength limit, which reads
\be
\delta{I}_s= \bar{s} \frac{\sigma (\delta \mathcal{E})^2 d^3 }{3 \eta c}\sgn{k}.
\ee

It is of interest to investigate the modification of these results at finite but small particle density, where $\bar{n} \ll \bar{s}$. In this regime, the intrinsic thermoelectric coefficient remains small, $\gamma/T \sim \bar{n}/\bar{s} \ll 1$, and can therefore be neglected. For an electron liquid driven purely by the wave, $\bar{\mathcal{E}} = 0$, the density and entropy perturbations become
\begin{subequations}\label{dndsE0}
\begin{align}
\delta n & = \frac{k \sigma  \delta \mathcal{E}}{e \omega} + \frac{\mathscr{G}(Kd)d^{2} k \delta \mathcal{E} (\omega \mathit{c_p} + i \kappa k^2) e\bar{n}^2}{ \omega \left(i\mathscr{G}(Kd) T \,d^{2} k^{2} \bar{s}^{2}+3 \,ik^{2} \eta  \kappa +3 \mathit{c_p} \eta \omega \right)}, \\
\delta s & = \frac{\mathscr{G}(Kd)d^{2} k \delta \mathcal{E}  \mathit{c_p} e\bar{s}\bar{n}}{ i\mathscr{G}(Kd) T \,d^{2} k^{2} \bar{s}^{2}+3 \,ik^{2} \eta  \kappa +3 \mathit{c_p} \eta \omega }
\end{align}
\end{subequations}
by solution of the conservation laws \rr{cons1}, cf. Appendix \ref{sec: perturbations}.
Note that the perturbations \rr{dndsE0} are odd with respect to $k$ and lead to an electric current \rr{dIe} with the same odd in $k$ parity
\begin{align}
\delta I_e  = &  \frac{e^2 \bar{n} \delta \mathcal{E} d^3}{3\eta}\Re \left\{ \delta n \left[ 1 +  \mathscr{G}^*(Kd) \right] \right\} \nonumber \\
& + \frac{ek \frac{T}{c_p} \bar{s} d^3}{3\eta}
\Re \left\{ i \delta n \delta s^* \mathscr{G}^*(Kd) \right\}.
\label{Ieodd}
\end{align}
The current we obtain represents a hydrodynamic analogue of the acoustogalvanic effect, namely the generation of a DC electrical current in a conductor induced by the propagation of an acoustic wave in the absence of an applied voltage. The distinction lies in the microscopic origin of the effect. In the conventional acoustoelectric effect, the current arises from the direct momentum transfer from the wave to the electrons. In the hydrodynamic mechanism considered here, it instead originates from the rectified drag force exerted on the electron liquid through density modulations induced by the wave.

In the long wavelength limit the expression for the current simplifies to
\be \label{Ieoddser}
\delta I_e = 2 e\bar{n} \frac{\sigma (\delta \mathcal{E})^2 d^3 }{3 \eta c}
\left[ 1 + \left(\frac{\bar{n}}{\Gamma} \right)^2\right]\sgn{k} + O(k^3),
\ee
where
\be \label{Gamma}
 \Gamma^2 = \frac{3\sigma \eta }{ (ed)^2}.
\ee
The entropy current \rr{dIs}
\begin{align}
\delta I_s  = &  \frac{e  \delta \mathcal{E} d^3}{3\eta}\Re \left\{ \bar{s}\delta n  + \bar{n} \delta s \mathscr{G}^*(Kd) \right\} \nonumber \\
& + \frac{k \frac{T}{c_p} \bar{s} d^3}{3\eta}|\delta s|^2
\Re \left\{ i  \mathscr{G}^*(Kd) \right\}
\label{Isodd}
\end{align}
is also odd with respect to $k$, whose asymptotic form in the long wavelength becomes
\be \label{Isoddser}
\delta I_s = \bar{s} \frac{\sigma (\delta \mathcal{E})^2 d^3 }{3 \eta c}\left[ 1 + 2\left(\frac{\bar{n}}{\Gamma} \right)^2\right]\sgn{k} + O(k^3).
\ee
Since $\delta I_e\to0$ as $\bar{n}\to0$ while $\delta I_s\neq0$, this raises the possibility of heat transfer in the absence of net charge transfer \cite{Andreev2022}.

\begin{figure}
\vspace{5pt}
\begin{center}
\includegraphics[height=2.6in,width=3.6in]{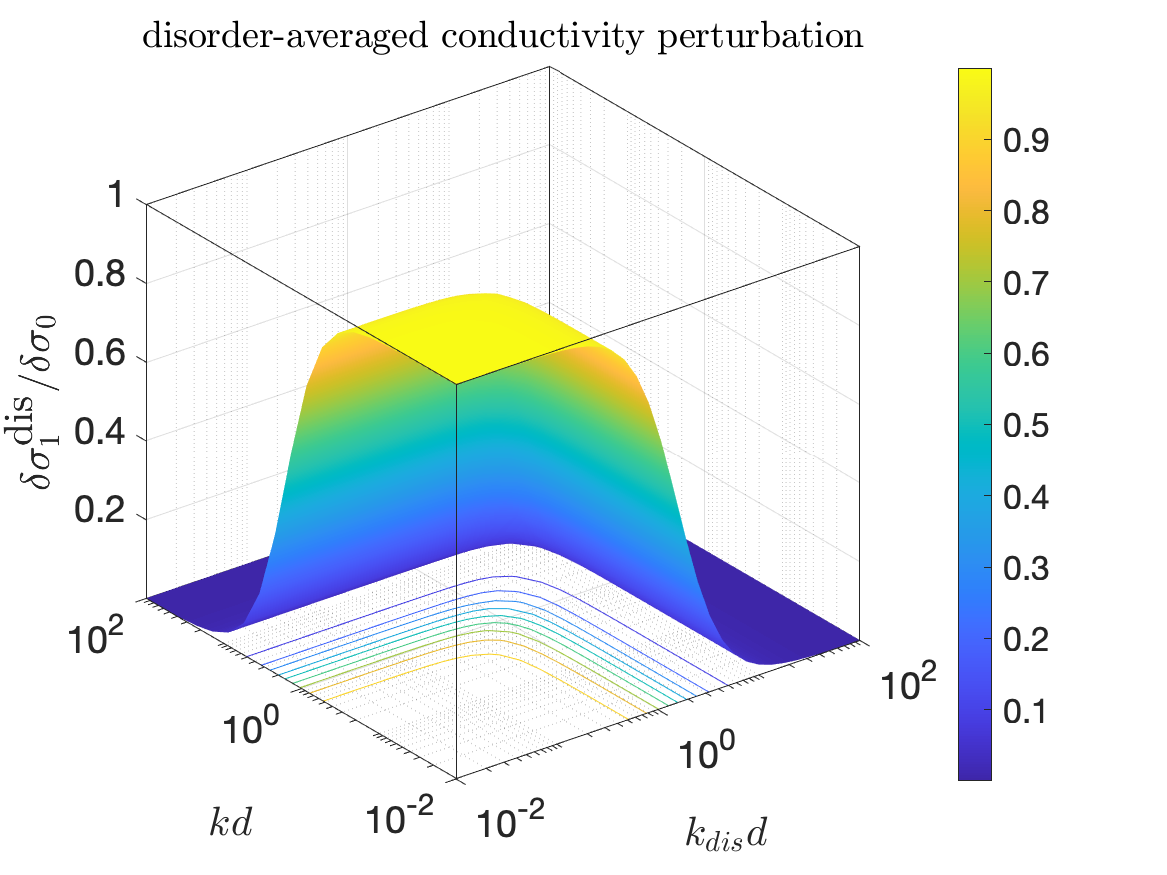}
\vspace{-0pt}
\end{center}
\caption{Disorder-averaged wave-induced conductivity enhancement, normalized by the long-wavelength limit conductivity \rr{dsigma}. Here, $k_{\text{dis}}$ denotes the disorder wavenumber, $k$ the excitation wavenumber, and $d$ the half-width of the Hall bar. As the disorder strength increases, $k_{\text{dis}} d > 1$, the conductivity enhancement induced by the wave is progressively suppressed.
\label{deltasigma_dis}}
\vspace{-0pt}
\end{figure}

\section{\label{sec: disorder}Weakly disordered systems}

In this section we consider the effect of weak disorder on the wave-induced entrainment of charge and heat current in the hydrodynamic regime. Phenomenologically,  disorder-induced momentum relaxation may be described by adding a friction force $-\varkappa \bm{u}$ to the Navier-Stokes equation \eqref{NS}.
When the correlation radius of the disorder potential exceeds the electron–electron relaxation length, the values of the friction coefficient $\varkappa$ may be obtained by averaging the hydrodynamic equations over disorder realizations \cite{Lucas2016,Li2020}.
For example, in the model of charge-puddle disorder relevant to graphene devices \cite{Martin2008,Zhang2009}, the disorder coefficient was obtained in Ref.~\cite{Li2020} and, near charge neutrality, takes the form
\begin{equation}
\label{kappa-puddle}
\varkappa = \frac{e^2}{2\sigma}\langle n_{\rm dis}^2(\mathbf{r}) \rangle ,
\end{equation}
where $n_{\rm dis}(\mathbf{r})$ denotes the static density variations induced by the disorder potential (to be distinguished from the wave-induced amplitude $\delta n$), and the angular brackets indicate a spatial average over the system. Equation~\eqref{kappa-puddle} is the leading term in $\bar{n}/\bar{s}\ll1$; the general expression at finite density is given in Ref.~\cite{Li2020}. Since all results of this section are derived to the same order in $\bar{n}/\bar{s}$ (the thermoelectric coefficient and the difference between $c_p$ and $c_v$ are neglected at that order), Eq.~\eqref{kappa-puddle} is sufficient here.
As a result, all expressions for the currents of charge and heat presented in the previous section can be straightforwardly generalized to the disordered case, which amounts to the substitution (see Appendix \ref{sec: appdis} for details)
\begin{equation}
K \to K_{\text{dis}} = \sqrt{\frac{i\omega}{\nu} - k^2 - k^2_{\text{dis}}}, \quad k_{\text{dis}}=l^{-1}_G=\sqrt{\varkappa/\eta}.
\end{equation}
The comparison of the viscous and disorder friction forces gives a characteristic length scale $\sqrt{\eta/\varkappa}$,
that can be interpreted as an analogue of the Gurzhi length, $l_G$ \cite{Gurzhi1968}. It determines the crossover from the Poiseuille profile in the clean limit to the profile governed by the Navier–Stokes equation \rr{NS} in the disordered limit. In principle, $\varkappa$ also depends on the oscillation frequency $\omega$. However, in the low-frequency limit, the wavenumber retains its constant form to leading order in $\omega$. Therefore, disorder effectively cuts off the low-frequency/long-wavelength behavior of the response currents, as shown in Fig.~\ref{deltasigma_dis}.

For a traveling wave of sufficiently long wave length when $k_{\text{dis}}\gg\text{max}\{k,\sqrt{ck/\nu}\}$, the crossover is described by a single-parameter function
\begin{equation}
g(x)=\frac{3}{x^3}[x-\tanh x].
\end{equation}
For example, for the wave-induced conductivity we find
\begin{equation}
\label{Iedis}
\delta\sigma^{\textrm{dis}}=
\frac{\sigma^{2}(\delta \mathcal{E})^{2}d^2}{6\eta  c^{2}} g(k_{\text{dis}}d),
\end{equation}
which generalizes the previous result obtained in the clean limit. All other expressions for $\delta I_e$ and $\delta I_s$ can be generalized similarly taking into account the crossover function, specifically for $\bar{\mathcal{E}}\to0$, one finds
\begin{subequations}\label{IeIs-dis}
\be \label{Ieoddserdis}
\delta I_e^{\text{dis}}= 2 e\bar{n} \frac{\sigma (\delta \mathcal{E})^2 d^3 }{3 \eta c} g(k_{\text{dis}}d)\sgn{k},
\ee
\be
\label{Isoddserdis}
\delta I_s^{\text{dis}}=\bar{s} \frac{\sigma (\delta \mathcal{E})^2 d^3 }{3 \eta c}g(k_{\text{dis}}d)\sgn{k}.
\ee
\end{subequations}
The corresponding heat current $I_q$ is related to the entropy current $I_s$ by
$I_q=TI_s$.

\section{Summary}

In summary, we have developed a theory of electromotive entrainment of charge and heat currents by a traveling surface acoustic wave \rr{E} in graphene devices operating in the hydrodynamic regime. The nonlinear coupling responsible for entrainment originates from wave-induced charge-density modulations rather than from the inertia of the electron liquid. The resulting currents are proportional to the intrinsic conductivity of the electron liquid.

Our consideration focused on devices in the Hall-bar geometry. At charge neutrality, the electrical conductivity of the system is enhanced by the wave. For clean systems the enhancement is given by Eq. \rr{dsigma}. The enhancement persists in presence of weak disorder and is described by Eq. \eqref{Iedis}.

In the absence of a DC bias, the transport electric current vanishes at charge neutrality, whereas the entropy current remains finite, cf. Eqs. \rr{Isoddser} and \rr{Isoddserdis}. This behavior provides a mechanism for heat entrainment in an electron liquid without net charge transport. At small but nonzero particle density, a nonvanishing DC electric current in the direction of the wave travel arises in the absence of a DC bias. This current is quadratic in the amplitude of the traveling-wave field and odd in its wavenumber, cf. Eqs. \rr{Ieoddser} and \eqref{Ieoddserdis} derived for clean and disordered limits respectively.

The results apply in the hydrodynamic window of graphene devices near charge neutrality: temperatures $T\sim50$--$150$~K, where electron-electron collisions dominate over phonon and impurity scattering; channel half-widths $d$ large compared with the thermal length $\lambda_T=\hbar v/T\simeq0.1\,\mu$m at $100$~K but smaller than the momentum-relaxation (Gurzhi) length $l_G=\sqrt{\eta/\varkappa}$, i.e.\ $d\sim1$--$5\,\mu$m in encapsulated devices; densities $|\bar{n}|\ll\bar{s}\sim(T/\hbar v)^2\sim10^{11}$~cm$^{-2}$, i.e.\ $|\bar{n}|\lesssim10^{10}$~cm$^{-2}$, yet above the residual charge-puddle scale $\sim10^{9}$~cm$^{-2}$ of clean samples; and wave frequencies $\omega\ll\tau_{\rm ee}^{-1}\sim T/\hbar\sim10^{13}$~s$^{-1}$, comfortably satisfied by SAW in the GHz range, with wavelengths $2\pi/k\gtrsim d$ so that $kd\lesssim1$ and the width-averaged description of Appendix~\ref{sec: perturbations} applies. In gated devices the wave potential is strongly screened on the time scale of the SAW period ($\gamma_M\gg\omega$, see Appendix~\ref{sec: validity}); the amplitude $\delta\mathcal{E}$ entering our results is then the screened EMF, $|\delta\mathcal{E}|\simeq(\omega/\gamma_M)k|U|/e$.

\section*{Acknowledgements}

The work of A. L. was supported by NSF Grant No. DMR-2452658 and H. I. Romnes Faculty Fellowship provided by the University of Wisconsin-Madison Office of the Vice Chancellor for Research and Graduate Education with funding from the Wisconsin Alumni Research Foundation. The work of A. V. A. was supported by the National Science Foundation (NSF) Grant No. DMR-2424364.


\section*{Data availability}

All data presented in the figures were generated from analytical expressions derived and defined in the paper. The computer code used to produce the plots will be made available by the authors upon reasonable request.


\appendix

\section{Velocity fields induced by the traveling wave EMF}\label{app:velocity}

The body force $e n(x,t) \mathcal{E}(x,t)$ in the momentum equation \rr{NS}
is of the form
\begin{align} \label{enE0}
&e n(x,t) \mathcal{E}(x,t)= e\bar{n} \bar{\mathcal{E}} + \left[  \frac{1}{4}e\delta n \delta\mathcal{E}+\right. \nonumber \\
&\left.\frac{1}{2}e \left(\bar{n} \delta \mathcal{E} + {\delta n \bar{\mathcal{E}} }\right) e^{i(kx-\omega t)} + \frac{1}{4}e\delta n \delta\mathcal{E} e^{2i(kx-\omega t)}  + c.c. \right].
\end{align}
Near charge neutrality the thermodynamic state of the liquid is determined by the temperature alone, so that $\partial_xT = \frac{T}{c_p} \partial_x s$ with $c_p$ the heat capacity per unit area; this leads to
\begin{align} \label{sdxT}
&s\partial_xT  = s\partial_s T \partial_x s = \frac{sT}{c_p} \partial_x s \nonumber  \\
&=  ik\frac{T}{c_p} \left[ \frac{1}{2}   \bar{s} \delta s e^{i(kx-\omega t)} +\frac{1}{4} (\delta s)^2 e^{2i(kx-\omega t)}  \right]+c.c.  .
\end{align}
It is thus clear that the hydrodynamic velocity obtains the form \rr{u}.
The Navier-Stokes equations \rr{NS} become
\begin{subequations} \label{NS4}
\begin{align}
&\eta \partial_y^2 \bar{u}(y) + e\bar{n} \bar{\mathcal{E}}  =0, \\
&\eta \partial_y^2 u_0(y) +\frac{e\delta n \delta \mathcal{E}}{2} =0, \\
&{-i \rho \omega  u_1(y)}= \eta \partial_{[k]}^2 u_1(y) + e \left(\bar{n} \delta \mathcal{E} +  \delta n \bar{\mathcal{E}} \right) {- i k\frac{T}{c_p} \bar{s} \delta s},
\label{u1eq}
 \\
&{-2i \rho \omega  u_2(y) }= \eta \partial_{[2k]}^2u_2(y) + \frac{1}{2} e \delta n\delta \mathcal{E}{-\frac{1}{2} ik\frac{T}{c_p} (\delta s)^2}.
\label{u2eq}
\end{align}
\end{subequations}
where we introduced short-hand notations of the differential operators $\partial_{[k]}^2=\partial_y^2 -k^2$ and $\partial_{[2k]}^2=\partial_y^2 -(2k)^2$.

Averaging over the width $2d$ of the Hall-bar we obtain

\begin{subequations}\label{uav}
\begin{align}
&\langle \bar{u} \rangle=  \frac{e\bar{n} \bar{\mathcal{E}} d^2}{3 \eta}, \\
&\langle u_0 \rangle=  \frac{e\delta n \delta \mathcal{E} d^2}{6 \eta }, \\
&\langle u_1 \rangle = \frac{\left[ e \left(\bar{n} \delta \mathcal{E} + \delta n \bar{\mathcal{E}} \right) - i k\frac{T}{c_p} \bar{s} \delta s\right] d^2}{3\eta} \mathscr{G}(Kd), \\
&\langle u_2 \rangle = \frac{\left[ e\delta n \delta \mathcal{E} - ik\frac{T}{c_p} (\delta s)^2\right] d^2 }{6\eta  }  \mathscr{G}( K_2d),
\end{align}
\end{subequations}
where $K = \sqrt{\frac{i\omega}{\nu} - k^2}$, $K_2 = \sqrt{\frac{2i\omega}{\nu} - (2k)^2}$ and $\displaystyle \mathscr{G}(z )=\frac{3}{z^3}\left[\tan z - z \right] $.
The zero mode $u_0$ is a consequence of the quadratically nonlinear electric body force in \rr{NS}, cf. \cite{Shrestha2025,*Shrestha2025b,*Kirkinis2014zero}.

\begin{figure*}
\vspace{5pt}
\begin{center}
\includegraphics[height=2.2in,width=7.5in]{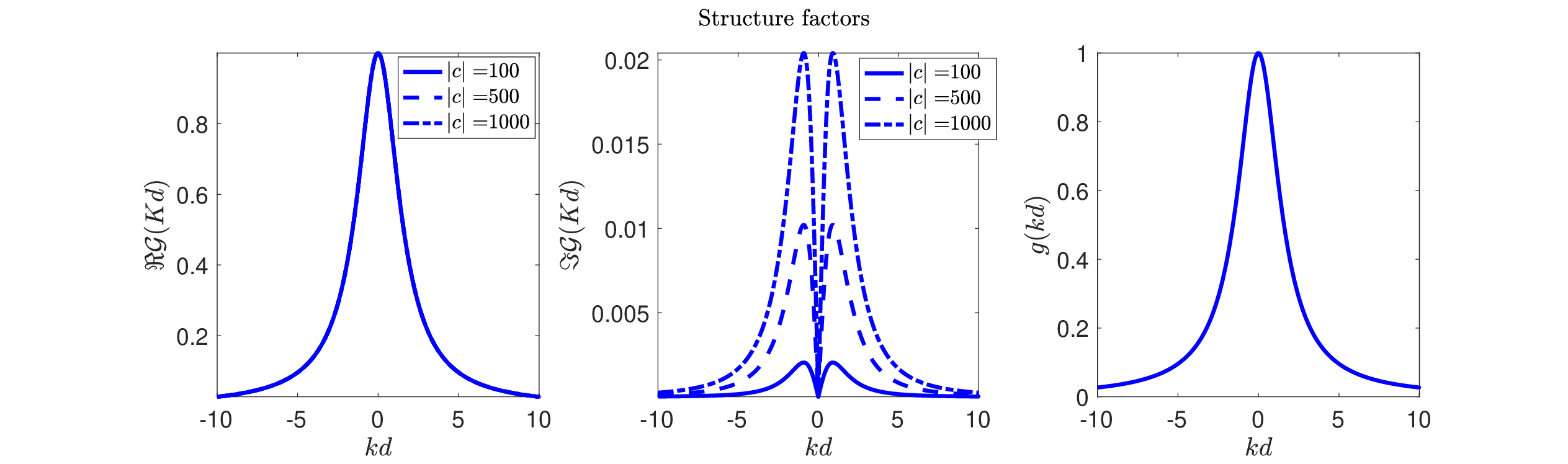}
\vspace{-0pt}
\end{center}
\caption{Structure factor functions $\mathscr{G}(kd)$ and $g(kd)$; see their asymptotic forms in Eqs.~\rr{ReGas}–\rr{gas}.
\label{structure_factor}}
\vspace{-0pt}
\end{figure*}

\section{\label{sec: structure}Structure factors}

With $K=\sqrt{\frac{ick}{\nu}-k^2}$ in terms of $c$ (fixed) and $k$ we expand  the real and imaginary parts of $\mathscr{G}(Kd) = \frac{3}{(Kd)^3}\left[\tan Kd - Kd \right] $ and that of $g(kd) = 3\frac{kd - \tanh kd}{(kd)^3} $ with respect to $k$. Thus,
\begin{subequations}
\begin{align}\label{ReGas}
\Re \mathscr{G}(Kd)& =\left\{
\begin{array}{ll}
1-\left[\frac{2}{5}+ \frac{17}{105} \left(\frac{cd}{\nu} \right)^2     \right](kd)^2
,& kd\ll1, \\
\frac{3}{(kd)^2} \left[ 1 - \frac{1}{kd} - \frac{c^2}{\nu^2} \frac{1}{k^2}\right]
,& kd\gg1,
\end{array}
\right.
\\
\label{ImGas}
\Im \mathscr{G}(Kd) & =\left\{
\begin{array}{ll} \frac{2cd^2}{5\nu} k \left[
1- (kd)^2 \left(\frac{17}{21}+ \frac{31 (dc)^2}{189\nu^2}    \right)  \right]
,& kd\ll1, \\
\frac{3c}{\nu k^3d^2} \left[ 1 - \frac{3}{2kd} - \frac{c^2}{\nu^2} \frac{1}{k^2}\right]
,& kd\gg1,
\end{array}
\right.\\
\label{gas}
g(kd)
&= \left\{
\begin{array}{cc}
1,& kd\ll1,\\
\frac{3}{(kd)^2} ,& kd\gg1.
\end{array}
\right.
\end{align}
\end{subequations}
The ubiquitous function $g$ is positive and is essentially a scaled Langevin function since $g(z) \equiv \frac{3 \tanh z}{z^2} \mathcal{L}(z)$, where $\mathcal{L}(z) = \coth z - \frac{1}{z}$.

\section{\label{sec: perturbations}Density and entropy perturbations}

We substitute the expansions \rr{ns} and \rr{u} into the charge and energy conservation laws \rr{cons1}, retain only terms of linear order (i.e. $\delta n$, $\delta s$ etc.), replace the temperature gradient $\partial_xT = \frac{T}{c_p} \partial_x s$, we average over the width of the Hall-bar,
and obtain
\begin{subequations}
\begin{align}\label{cdn}
\omega \delta n=& k\left[ \bar n u_1 + \delta n \bar{u} + \frac{\sigma \delta \mathcal{E}}{e}  - \frac{\gamma}{c_p} ik \delta s \right] \\
\omega \delta s = &k\left[ \bar{s} u_1 + \delta s \bar{u} - \frac{\kappa}{c_p} ik \delta s + \frac{\gamma}{\bar{T}} e \delta \mathcal{E} \right]
\label{cds}
\end{align}
\end{subequations}
Employing the average velocities \rr{uav}, we express \rr{cdn}, \rr{cds} as a system for the two unknowns $\delta n$ and $\delta s$
\begin{widetext}
\be
\label{dndssys}
\left( \begin{array}{cc}
-\omega +\frac{k e \bar{n}  \,d^{2} \left( \mathscr{G}(Kd) +1\right)  \bar{\mathcal{E}} }{3 \eta}& \frac{-i k^{2} \left( \mathscr{G}(Kd) T \,d^{2} \bar{n}  \bar{s} +3 \eta  \gamma \right)}{3\mathit{c_p} \eta} \\
\frac{k \bar{s} e  \bar{\mathcal{E}}  \,d^{2}  \mathscr{G}(Kd)}{3 \eta}  &    -\omega +\frac{k e \bar{n}  \bar{\mathcal{E}} \,d^{2}}{3 \eta}-\frac{ik^2\left( \mathscr{G}(Kd)T \,d^{2} \bar{s}^{2}+3 \eta  \kappa \right) }{3 \mathit{c_p} \eta}
\end{array}
\right)
\left( \begin{array}{c}
\delta n \\ \delta s
\end{array}
\right) = -
\left( \begin{array}{c}
\frac{k \mathit{\delta \mathcal{E} } \left( \mathscr{G}(Kd)d^{2} e^{2} \bar{n} ^{2}+3 \sigma  \eta \right)}{3 \eta  e}
 \\ \frac{\mathit{\delta \mathcal{E} } e \left(\bar{n}  \bar{s} T d^{2}  \mathscr{G}(Kd) +3 \eta  \gamma \right) k}{3 \eta  T}
\end{array}
\right)
\ee
\end{widetext}
It can be readily verified that for $\gamma/T\sim \bar{n}/\bar{s}$ near charge neutrality, the contributions to density variations containing the intrinsic thermoelectric coefficient can be neglected if $\mathcal{G}(Kd) d^2\bar{s}^2\gg 3\eta$. Indeed, in the long-wavelength limit, the structure factor $\mathcal{G}$ is of order unity; therefore, this condition can be equivalently restated as $d\gg\lambda_T$. Here we have used the estimates $\eta\sim \bar{s}\sim 1/\lambda^2_T$ with $\lambda_T\sim v/T$ denoting the thermal de Broglie length. The condition $d\gg\lambda_T$ is, in essence, the applicability criterion of the hydrodynamic regime.

\section{\label{sec: currents}Clean case currents}

The (DC) electric current can be written in the form
\begin{subequations}\label{Ie}
\begin{align}
&I_e = \bar{I}_e+ \delta I_e, \\
&\bar{I}_e \equiv  \int_{-d}^d
\left[e\bar{n} \bar{u}(y) + \sigma \bar{\mathcal{E}} \right] dy, \\
&\delta I_e =\Re  \int_{-d}^d e \left\{ \bar{n}u_0(y) + \frac{\delta n^*}{2} u_1(y)  \right\} dy
\end{align}
\end{subequations}

$  \bar{I}_e$ is the net linear current as determined in Ref. \cite{Li2022} and $\delta I_e$ is the nonlinear perturbation induced by the traveling wave field.
The net linear current becomes
\be
\bar{I}_e = 2d \sigma_e \bar{\mathcal{E}},
\ee
where $\sigma_e = \sigma  + \frac{(e\bar{n}d)^2}{3\eta}
$ is the effective linear conductivity.

The entropy current is of the form
\begin{subequations} \label{Is}
\begin{align}
& I_s = \bar{I}_s+ \delta I_s, \\
&\bar{I}_s \equiv  \int_{-d}^d
\left[\bar{s} \bar{u}(y) +  \frac{\gamma}{T} e\mathcal{E}\right] dy, \\
&\delta I_s =\Re  \int_{-d}^d  \left\{ \bar{s}u_0(y) + \frac{\delta s^*}{2} u_1(y)  \right\} dy
\end{align}
\end{subequations}
and is likewise
composed of a linear term and a perturbation. The linear term is
\be \label{barIs}
\bar{I}_s =2d\left[ \frac{d^2}{3\eta} e\bar{n} \bar{s} +  \frac{\gamma}{T} e \right]\bar{\mathcal{E}}.
\ee

\section{\label{sec: appdis}Weak disorder}

We solve the Navier-Stokes equations with the disorder-induced friction force as described in section \ref{sec: disorder}. The average velocities over the width of the channel become
\begin{subequations} \label{uavdis}
\begin{align}
&\langle \bar{u} \rangle=  \frac{e\bar{n} \bar{\mathcal{E}} d^2}{3 \eta}g(k_{\text{dis}} d), \\
&\langle u_0 \rangle=  \frac{e\delta n \delta \mathcal{E} d^2}{6 \eta }g(k_{\text{dis}} d), \\
&\langle u_1 \rangle = \frac{\left[ e \left(\bar{n} \delta \mathcal{E} + \delta n \bar{\mathcal{E}} \right) - i k\frac{T}{c_p} \bar{s} \delta s\right] d^2}{3\eta} \mathscr{G}(K_{\text{dis}}d), \\
&\langle u_2 \rangle = \frac{\left[ e\delta n \delta \mathcal{E} - ik\frac{T}{c_p} (\delta s)^2\right] d^2 }{6\eta  }  \mathscr{G}( K_{2\text{dis}}d),
\end{align}
\end{subequations}
where $K_{\textrm{dis}} = \sqrt{\frac{i\omega}{\nu} - k^2-k^2_{\textrm{dis}}}$, $K_{2\textrm{dis}} = \sqrt{\frac{2i\omega}{\nu} - (2k)^2-k^2_{\textrm{dis}}}$,
$\displaystyle \mathscr{G}(z )=\frac{3}{z^3}\left[\tan z - z \right] $ and $g(x) = 3\frac{x - \tanh x}{x^3}$.

\section{\label{sec: validity}Self-consistency and validity of approximations}

In this section, we analyze additional contributions arising from the self-consistent emf and we \emph{a-posteriori} validate the assumptions made in Eq. \rr{NS} and \rr{E}.
The induced density variations by a traveling wave lead to an additional electric field
\begin{equation}
\delta\bm{E}=-e\mathbf{\nabla}\int d^2\mathbf{r}'\frac{\delta n(\mathbf{r}')}{|\mathbf{r}-\mathbf{r}'|}
\end{equation}
that can be obtained from the Poisson equation. This relation simplifies in gated structures, and takes the spatially local form
\begin{equation}
\label{Eind-local}
\delta\bm{E}=-\frac{e}{C}\mathbf{\nabla}\delta n,
\end{equation}
where $C=\varepsilon/4\pi d'$ is the gate-to-channel capacitance per unit area, $d'$ is the distance to the gate, and $\varepsilon$ is the dielectric constant.
This approximation neglects the long-ranged (dipole-type) part of the Coulomb interaction (screened by the gate) and is valid as long as the charge density $\delta n$ varies on length scales much longer than $d'$.
By considering this limit, we are guided by recent experimental results \cite{Domaretskiy2025}, which demonstrate that proximity screening greatly enhances the electronic quality of graphene{\color{blue},} an effect that is particularly important for realizing the hydrodynamic regime.

The induced field~\eqref{Eind-local} enters the problem in two ways. At linear order it is part of the EMF~\eqref{E}: writing $e\mathcal{E}=-\partial_x(U+\mu+e\phi)$ with $\mu+e\phi=(e^2/C)\delta n$ and using Eq.~\eqref{dnds}, one finds 
\begin{equation}
\delta\mathcal{E}=-\frac{ikU}{e}\,\frac{1}{1+i\gamma_M/\omega},\qquad \gamma_M=\frac{\sigma k^2}{C}=\frac{4\pi\sigma k^2d'}{\varepsilon},
\end{equation}
where $\gamma_M$ is the gate-screened Maxwell relaxation rate of the charge modulation; this fixes the coefficient $\alpha$ introduced after Eq.~\eqref{ns}, and all results above, being expressed through $\delta\mathcal{E}$, hold for arbitrary $\gamma_M/\omega$. For $\gamma_M\gg\omega$, which is the typical situation for SAW frequencies in gated devices, the density modulation follows the wave adiabatically, $|\delta n|\simeq C|U|/e^2$, and $|\delta\mathcal{E}|\simeq(\omega/\gamma_M)k|U|/e$. At quadratic order the induced field produces the force density $(e^2/C)\delta n\nabla\delta n=\nabla(e^2\delta n^2/2C)$, a total derivative whose average over the period vanishes; it therefore does not contribute to the DC currents \eqref{dIe-def} and \eqref{dIs} and only affects the second harmonic $u_2$, which drops out of them.

\bibliography{biblio-SAW}

\end{document}